# Exploring the World of Rhamnolipids: A critical review of their production, interfacial properties, and potential application

Eduardo Guzmán,[1,2,*] Francisco Ortega,[1,2] Ramón G. Rubio[1,2]

[1] Departamento de Química Física, Facultad de Ciencias Químicas, Universidad Complutense de Madrid. Ciudad Universitaria s/n. 28040-Madrid (Spain)

[2] Instituto Pluridisciplinar, Universidad Complutense de Madrid. Paseo de Juan XXIII 1. 28040-Madrid (Spain)

[*] To whom correspondence should be addressed: eduardogs@quim.ucm.es (Professor Eduardo Guzmán)

**Abstract**

Rhamnolipids are very promising sugar-based biosurfactants, generally produced by bacteria, with a wide range of properties that can be exploited at an industrial and technological level, e.g., in cosmetics, food science or oil recovery, to provide benefits for human health and the environment. This has led to intensive research into optimising their production to increase yields and minimise costs, which is challenging because biotechnological methods for rhamnolipid production result in complex product mixtures and require the introduction of complex strategies to ensure the purity of the rhamnolipid obtained. This is an important issue for the introduction of rhamnolipids to the market due to the differences that exist between the properties of the different congeners. This review attempts to provide an overview of the interfacial properties, potential applications, and recent advances in understanding the molecular mechanisms that govern the adsorption to interfaces and assembly in solution of rhamnolipids. In addition, the review also discusses some general aspects related to the production and purification methods of rhamnolipids, highlighting the need for further research to fully exploit their potential. It is hoped that this review will contribute to the growing body of knowledge about rhamnolipids and stimulate further research in this field.

**Introduction**

Surfactants are amphiphilic molecules characterized by their ability to reduce the interfacial tension between immiscible fluids, such as oil and water. This has stimulated their use to control a wide range of processes of industrial relevance, such as detergency, emulsification, and foaming [1]. However, current problems related to health, pollution, climate change and depletion of fossil fuel reserves require solutions to the massive use of conventional surfactants, generally derived from petrochemical sources [2]**. In this context, biosurfactants and bio-based surfactants have emerged as promising candidates due to their unique properties and potential applications [3, 4]*. The former are surfactants derived directly from natural renewable sources, usually microorganisms through their metabolic pathways, and produced industrially by fermentation processes, while the latter are derived directly from renewable feedstocks by chemical or enzymatic processes involving the modification of bio-based feedstocks, but do not need to be produced directly by living organisms [5]**.

Biosurfactants are a diverse group of surface-active molecules produced mainly, but not exclusively, by microorganisms (bacteria, fungi, or yeast) as secondary metabolites, either on the cell surface or secreted extracellularly, and involved in many of the cellular communication processes [6]. In general, biosurfactants appear as a thin film on the surface of the producing microorganisms and contribute to their detachment or attachment to other cell surfaces [3]. Among the most common microorganisms that produce biosurfactants are several examples belonging to the genera *Pseudomonas*, *Bacillus*, *Candida*, *Rhodococcus* and *Corynebacterium* [7].

Biosurfactants are in most cases of anionic or non-ionic nature and can be divided into two groups according to their molecular weight: (i) low molecular mass (LMM) surfactants (molecular weight in the range <1200 g/mol) and (ii) high molecular mass (HMM) surfactants (molecular weight >45,000 g/mol) [8]. The former are effective in reducing interfacial tension, while HMM are more effective in stabilizing emulsions and foams [9]. A more detailed classification of biosurfactants is usually made taking into account the differences in the chemical structure of the different families, which can be divided into two well-differentiated parts: (i) hydrophobic moiety (saturated/unsaturated fatty acids), and (ii) hydrophilic moiety (amino acids/peptides, anions/cations or mono-/di-/polysaccharides) [3, 10]. This is important because the diversity of chemical structures of biosurfactants, which can be tailored for specific applications through genetic engineering or manipulation of the producing microorganisms, results in different functional properties [11-17]. These properties make biosurfactants an attractive area of research with the potential to make significant contributions to various fields, such as bioremediation, food processing, cosmetics, and pharmaceuticals. Table 1 summarizes the most common families of existing biosurfactants according to their chemical structure. Biosurfactant classification can also be made according to their charge, molecular weight, or secretion type [18].

The use of biosurfactants in various fields takes advantage of the improved properties of biosurfactants compared to traditional synthetic surfactants. These include lower toxicity, higher biodegradability and benign interaction with the environment, higher foaming properties and tolerance to extreme conditions (high temperature and salinity, low and high pH). In addition, the critical micelle concentration (CMC) of biosurfactants is lower than that of synthetic surfactants, which helps to increase their effectiveness in a wide range of applications. On the other hand, several biosurfactants exhibit antibacterial, antifungal, antiviral or anti-tumor activity, which open important venues for their use in therapeutic applications [18, 19]. Despite the many benefits of biosurfactants, there are still a number of challenges and limitations to overcome. A key challenge is the relatively high production cost compared to synthetic surfactants, which may limit their commercial viability for certain applications [20, 21]. Further limitations arise from the lack of standardized methods for characterizing biosurfactants and assessing their performance, together with the heterogeneity of biosurfactant samples and the lack of surfactant expertise among those who produce biosurfactants [22]*. This makes it difficult to compare results between studies and determine their true potential [23, 24]. In addition, there are concerns about the potential ecological impacts associated with large-scale production of biosurfactants [25]. These limitations highlight the need for further research and development to fully unlock the potential of biosurfactants. For instance, in most cases surfactants are only one part of commercial formulations, so a proper comparison with synthetic surfactants requires an understanding of the synergistic behavior of the whole set of components in the formulation **[26, 27][28]**.

Table 1. Most common families of biosurfactants

| Family | Examples | References |
|---|---|---|
| Glycolipids | Rhamnolipids Sophorolipids mannosylerythritol lipids | [3, 10] |
| Lipopeptides | Surfactin Iturin Fengycin | [13, 14] |
| Fatty acid/Phospholipids/Neutral lipids | Phosphatidylcholine Phosphatidylethanolamine. | [15, 16] |
| Polymeric biosurfactants | Emulsan Liposan Lipomanan Alasan | [17, 28]** |

Glycolipids are probably the most studied low molecular weight biosurfactants and have attracted much attention due to their potential biotechnological applications, especially because of their versatility and their possible production from a wide range of renewable resources, including hydrocarbons, industrial wastes, frying and olive oil wastes, and agricultural by-products. From a chemical point of view, the general structure of glycolipids consists of one or more carbohydrate rings attached to one or more fatty acids by ether or ester linkages. The specific nature of the fatty acid and carbohydrate ring contained in a given glycolipid give rise to the diversity of glycolipids, with the nature of the carbohydrate ring providing the basis for its classification [10]. In fact, sophorolipids are characterized by the presence of a sophorose polar head (two glucose rings linked through a β-1,2 glycosidic bond). In rhamnolipids, rhamnose rings, one in mono-rhamnolipids or two in di-rhamnolipids, form the polar head. This is formed by 4-*O*-β-d-mannopyranosyl-meso-erythritol in the mannosylerythritol lipids. There are also other families of glycolipids, such as trehalose lipids, xylolipids and cellobiose lipids, which have been less studied [6]. Among the different families of glycolipids, rhamnolipids have been the subject of extensive research due to their high production yield after relatively short incubation times and their relatively high interfacial activity [29]**. The latter gives them applications such as oil tank cleaning or enhanced oil recovery in the petroleum industry, or the removal of contaminants (oils, pesticides, or heavy metals) from soil and water. In addition, their high interfacial activity gives them good foaming and wetting properties, which are important in the cosmetics, personal care, and food industries [30]. This has stimulated the growth of the rhamnolipid market , which is expected to increase to €2.6 billion by 2023 [31].

This review aims to provide a comprehensive overview of the current state of knowledge on the interfacial properties of a specific family of biosurfactants belonging to the group of the glycolipids, the rhamnolipids, and their potential applications, while critically evaluating the recent advances and identifying the gaps that need to be addressed in future research. In particular, the interfacial properties of rhamnolipids will be reviewed, with a special emphasis on their potential applications in various fields, such as enhanced oil recovery, food and cosmetic industries, pharmaceuticals, and environmental remediation. We will also discuss the recent advances in understanding the molecular mechanisms underlying the interfacial properties of rhamnolipids, including their self-assembly behavior, molecular packing at interfaces, and interactions with other molecules. Furthermore, we will highlight the current directions in the study of rhamnolipids as biosurfactants, including their production and purification.

**Structure and production of rhamnolipids**

Rhamnolipids are glycosides consisting of a polar moiety (glycon part) composed of one or two di-L-rhamnose units linked to each other by an α-1,2-glycosidic linkage, resulting in the formation of mono-rhanolipids and di-rhamnolipids, respectively. On the other hand, the hydrophobic moiety (aglycon part) consists of saturated or unsaturated (mono- or polyunsaturated) β-hydroxy fatty acid chains of variable length (between C8 and C24) linked together by an ester linkage. The glycon and aglycon parts are linked by a glycosidic linkage. The combination of different rhamnose units with different types of β-hydroxy fatty acids results in a number of congeners close to 60 [32], which are commonly produced by different organisms [33, 34]. Figure 1 shows the general molecular formula of mono- and di-rhamnolipids.

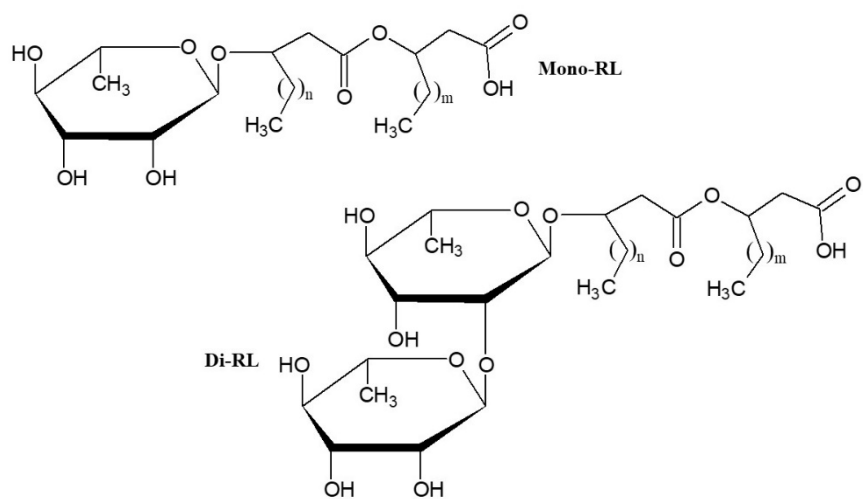

Figure 1. General molecular formula of mono- and di-rhamnolipidss. The subindexes m and n indicate the number of methyl groups in the alkyl chains of the rhamnolipids, it is possible to find rhamnolipids characterized by n=m, but also asymmetric ones with n≠m.

Rhamnolipids were first isolated from *Pseudomonas aeruginosa* (hereinafter *P. aeruginosa*) by Jarvis and Johnson [35]** more than 70 years ago. This seminal work stimulated extensive research to exploit the use of this opportunistic pathogen to produce rhamnolipids. This has led to the isolation of up to four different rhamnolipids using *P. aeruginosa* strains: 3-[3-(2-*O*-α-L-rhamnopyranosyl-α-L-rhamnopyranosyloxy)decanoyloxy]decanoic acid (di-rhamnolipid with two saturated C10 hydrophobic tails), 3-[(6-deoxy-α-L-mannopyranosyl)oxy]decanoic acid (mono-rhamnolipid with one C10 hydrophobic tail), 3-[3-(α-L-rhamnopyranosyloxy)decanoyloxy]decanoic acid (mono-rhamnolipid with two saturated C10 hydrophobic tails) and 3-[(2-*O*-α-L-rhamnopyranosyl-α-L-rhamnopyranosyl)oxy] decanoic acid (di-rhamnolipid with one C10 hydrophobic tail) [34, 36]. It is worth noting that although the Gram-negative bacterium *P. aeruginosa* can be considered as the most common microorganism producing rhamnolipid [18], other species of *Pseudomonas,* such as *Pseudomonas alcaligenes*, *Pseudomonas fluorescens*, *Pseudomonas chlororaphis*, *Pseudomonas putida*, or *Pseudomonas stutzeri,* can also be used in the production of rhamnolipids [37]. Unfortunately, most *Pseudomonas* bacteria are pathogenic microorganisms, which is often a problem for scaling up the production of rhamnolipids. In addition, the complex regulatory mechanism of rhamnolipid production by *Pseudomonas* is a challenge for their application in large-scale production. This is clear when one considers that the regulation of the rhamnolipid production involves a set of five proteins/enzymes (RhlA, RhlB, RhlC, RhlG and RhlI), together with the action of the *rhlABR* gene from *P. aeruginosa* which is responsible for the synthesis of the RhlR regulatory protein and of a rhamnosyl transferase, both of which are crucial for the synthesis of rhamnolipids [38]. It is therefore necessary to look for alternative bacterial species, preferably with non-pathogenic character, to facilitate production and broaden the range of applications. Such alternatives include *Burkholderia* species (*Burkholderia thailandensis*, *Burkholderia plantarii* or *Burkholderia pseudomallei*) [39, 40], different *Planococcus* species (*Planococcus halotolerans* and *Planococcus rifietoensis*) [41], *Lysinibacillus sphaericus* [42], *Enterobacter asburiae, Marinobacter* species [43], or *Acinetobacter calcoaceticus* [44]*.The latter allows the isolation of complex mixtures containing mainly mono- and di-rhamnolipids with decyl hydrophobic chains together with minor amounts of up to six other congeners. Indeed, the selection of a specific bacteria strain is a critical issue to produce rhamnolipids because it defines the obtained rhamnolipids, and the number of congeners [37, 45]. Therefore, the complex relationship between bacterial strains and the resulting rhamnolipid congeners underscores the importance of strain selection in ensuring specific product outcomes and purity. In fact, even within the same family of microorganisms, different strains can produce markedly different mixtures of rhamnolipids. This variance not only underscores the need for precise strain selection, but also highlights the impact of fermentation conditions on the composition of the final product. The slightest change in fermentation parameters can significantly affect the rhamnolipid profile, further complicating the quest for consistent product purity. In addition, maintaining uniformity in the analysis of different batches is essential to make meaningful comparisons between experimental studies of rhamnolipids or other biosurfactants. The adoption of standardized methodological approaches across studies is critical to ensure the reliability and accuracy of results. Discrepancies in the composition or purity of rhamnolipid mixtures used in different studies could inadvertently lead to contradictory results [46]**. Therefore, understanding the intricacies of rhamnolipid production as influenced by bacterial strains, fermentation conditions, and analytical methods is critical. Recognizing and accounting for these variables is essential to ensure consistency and reliability when comparing results from different studies in this field.

The high demand for biosurfactants has led to the development of biotechnological tools to identify the biological pathways involved in biosurfactant production, and to obtain hyperproducing strains or recombinant mutants [7]. The most common scheme for isolating biosurfactants from microbial strains is shown in Figure 2.

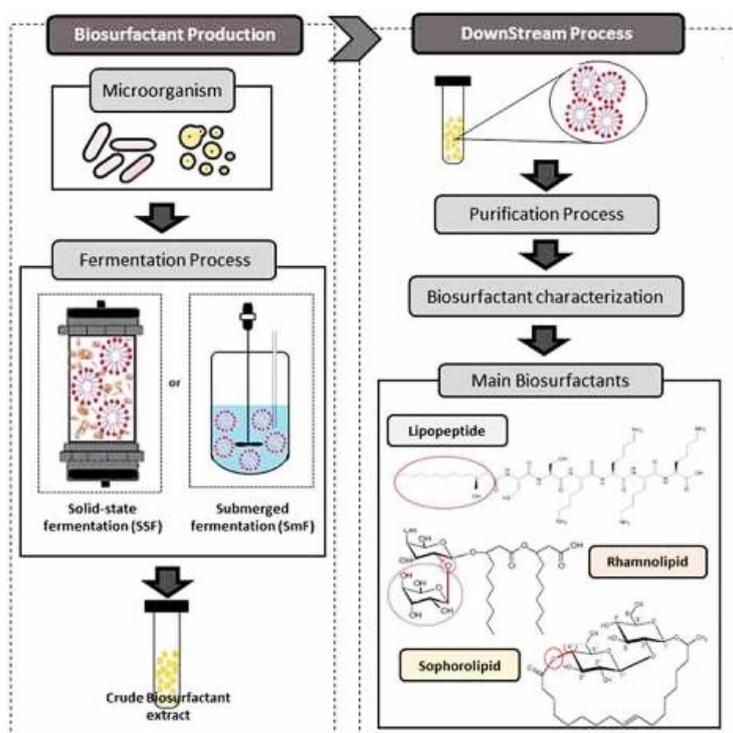

Figure 2. Sketch of the process of isolation, purification, and characterization of biosurfactants. Adapted from Eras-Muñoz et al. [7]. Licensed under Creative Common Attribution 4.0 International License.

The production of rhamnolipids is most commonly achieved by fermentation processes such as submerged fermentation (SmF) or solid-state fermentation. (SSF) [7], accompanied by separation steps to remove the products from the culture medium [47]\*\*. This separation usually involves sedimentation, centrifugation and/or extraction steps. However, there are problems that affect the yield of the production of rhamnolipids [48, 49]. The first is related to the inhibitory effect on the production of the rhamnolipids produced, and the second is related to the complexity of the separation and purification steps and their high energy consumption. In fact, the separation and purification processes represent approximately the 60% of the costs associated with the production of rhamnolipids [7]. On the other hand, the use of large quantities of organic solvents during the production process can cause pollution. It is therefore very important to carefully optimize several parameters, including the solubility of the carbon source, the type of feed, pH, temperature, aeration rate, dissolved oxygen, cell density and the ability to remove the product *in situ*, to ensure an efficient production of rhamnolipids [50]. In particular, the solubility of the carbon source in the culture media is of a paramount importance in controlling the production yield. The latter is critical because it is common that the production of rhamnolipids does not exceed of 10 g/L for fermentation processes [51-63], arriving up to values close to 100 g/L by using hyperproducer strains [3, 64-67].

Rhamnolipids can be produced using both water-soluble carbon sources, such as glycerol, glucose, mannitol or ethanol, and water-insoluble carbon sources, such as n-alkanes and olive oil. The type of carbon source influences the type of biosurfactant produced. However, the chain length of the substrates does not cause any variability in the chain length of the fatty acid moieties [68]. It is common that the highest production yields of rhamnolipids are obtained using oily carbon sources. However, the use of cheap and renewable carbon sources, such as lignocellulose or palm oil agricultural refinery waste, such as palm fatty acid distillate (PFAD), can help to reduce production costs [69, 70]. Together with the type of carbon sources, the supplementation of nitrogen in the medium is very important to produce biosurfactants in general, and rhamnolipids in particular are influenced by nitrogen supplementation. In the case of rhamnolipids, the source of nitrogen for the optimization of the production is in the form of nitrates. The optimal production of rhamnolipids can be achieved with a carbon to nitrogen ratio in the range of 18:1 to 16:1, whereas the production of rhamnolipids is inhibited with a carbon to nitrogen ratio of less than 11:1 [68]. Finally, the content of dissolved oxygen also allows controlling the production yield [71]. The substrate feeding profile also plays a significant role in the production of rhamnolipids as it controls microbial growth in the lag and growth phases. In addition, the pH of the culture media should be fixed at around 7, while remaining slightly acidic (6-6.5) during the stationary and death phases of fermentation [72]. It should be noted that maximizing the production of rhamnolipids requires obtaining a strain with a high yield. The most widely used method to obtain favorable strains is to isolate them from environments where biosurfactant-producing strains are abundant. Many laboratories have carried out extensive screening to isolate and categorize different strains that produce rhamnolipids. In particular, *P. aeruginosa* strains have consistently shown the highest yields of rhamnolipid production among the bacteria studied [73]**. Table 2 summarizes some examples of rhamnolipid-producing bacterial strains and the maximum rhamnolipid production yield obtained, together with the carbon source used for such production. It is worth noting that the comparison between the reported production yields of rhamnolipids in different studies is a significant challenge, mainly due to discrepancies in the units used to define these yields. Beyond simply understanding the production yield in terms of concentration, which could be quantified in mass-based units (such as weight percent or g/L) or number of molecules (mol/L), it is crucial to delve deeper into the specific composition of the rhamnolipid mixtures obtained. In fact, the characterization of rhamnolipid purity often involves quantifying the concentration of the desired congeners within the total rhamnolipid mixture. Analyzing the composition of congeners aids in evaluating their specific functionalities and understanding their role in various applications. On the other hand, when discussing production yields, the units used highlight different aspects of concentration. The choice of units depends on the specific focus of the analysis. For instance, mass-based units are often used to determine the concentration in terms of the amount of rhamnolipids produced, while mol-based units provide insights into the molecular concentration, crucial for understanding interactions in various applications. Unfortunately, obtaining detailed compositional data can be challenging. Often, reported yields refer to the total quantity of rhamnolipids obtained without specifying the individual components or their respective concentrations. The lack of detailed compositional information creates complexity when comparing production yields across studies. Researchers are unable to fully understand the precise nature and proportion of different rhamnolipid species produced without insight into the specific composition of the rhamnolipid mixtures. This gap hinders the standardization of comparisons and the ability to draw precise conclusions about rhamnolipid production efficiency. Tackling this challenge demands a collaborative endeavor within the scientific community to prioritize all-inclusive reporting of rhamnolipid structure along with the overall yield data. This will aid researchers in presenting more refined observations into the composition of these mixtures, thereby enabling a more precise and consequential comparison of rhamnolipid yield across various studies. This improved comprehension would not only enhance the dependability of comparative analyses but also promote a broader understanding of the rhamnolipid production processes and their potential applications [74].

Table 2. Examples of rhamnolipid-producing bacterial strains, and their maximum rhamnolipid production yield together with the carbon source used.

| Bacteria | Carbon source | Maximum yield (g/l) | References |
|---|---|---|---|
| *P. aeruginosa* FA1 | peanut meal | 2.6 | [61] |
| *P. aeruginosa* | canola oil | 3.2-3.6 | [51] |
| *P. aeruginosa* | frying oil+glucose | 4.2 | [52] |
| *P. aeruginosa* NY3 | frying oil | 9.1 | [57] |
| *P. aeruginosa* SG | glycerol | 21.5 | [67] |
| *P. aeruginosa* YM4 | glycerol / soybean oil | 24 / 25 | [65] |
| *P. aeruginosa* PAO1 | sunflower oil | 36.7 | [64] |
| *P. aeruginosa* LBI 2A1 | guava seed oil | 39.9 | [66] |
| *P. aeruginosa* O-2-2 | soybean oil | 70.6 | [72] |
| *Pseudomonas stutzeri* NCIM 5136 | glycerol | 4.7 | [53] |
| *Pseudomonas putida* KT2440 | glycerol | 3 | [55] |
| *Pseudomonas putida* | frying oil | 4.1 | [56] |
| *Pseudomonas taiwanensis* VLB120 | lignocellulosic biomass | 0.41 | [63] |
| *Achromobacter sp.* (PS1) | lignocellulosic hydrolysate | 5.5 | [54] |
| *Burkholderia thailandensis* E264 | glycerol | 4.6 | [58] |
| *Marinobacter* sp. MCTG107b | glucose | 0.74 | [59] |
| *Escherichia coli* (recombinant) | palm oil | 0.12 | [60] |
| *Bacillus subtilis* SPB1 | glucose | 5.7 | [62] |
| *Planococcus rifietoensis* IITR53 | glucose | 2.5 | [41] |
| *Planococcus halotolerans* IITR55 | | 1.8 | |

Bioengineering strategies can be viable alternatives to increase the production of rhamnolipids. Extensive research has been conducted to identify the genes that play a critical role in rhamnolipid production to obtain potentially safe and efficient rhamnolipid-producing strains [73]**. In fact, the use of genetic engineering or random mutagenesis has helped to increase the purity and production yield of rhamnolipids. The former approach focuses on obtaining a large number of bacterial strains involved in gene expression, while the random mutagenesis is based on the creation of fundamental changes in different bacterial strains, but it does not affect to biosynthetic enzymes or genes [29]**.

Currently, the research on rhamnolipids has focused on developing sustainable and cost-effective methods for their production, as the current production costs are significantly higher compared to petrochemical-based surfactants [75]*, and exploring their potential in various fields [18]. Some companies involved in the rhamnolipid production are Henkel, Lion Corporation, Allied Carbon Solutions Ltd., TeeGene Biotech, AGAE Technologies, LLC and Paradigm Biomedical Inc. [30]. It should be noted that although rhamnolipids have been known for about eight decades, their mass production has only recently become possible **[29, 76]. In fact, Evonik has recently brought rhamnolipids to the business-to-consumer (B2C) market. This has been achieved by using production methods that ensure a final product of appropriate quality, but do not require a high technical effort that increases production costs. In the case of Evonik, production started in 2018 through a subsidiary (Evonik Fermas) and is expected to reach 600 tons in 2020 at a total cost of €12.3 million [77]. In addition, a fully biodegradable commercial rhamnolipid (REWOFERM® RL 100) isolated from renewable raw materials was launched by Evonik at the end of 2022 [78].

At the industrial level, biotechnological approaches based on fermentation steps combined with the required downstream process are used. It is common to use non-pathogenic microorganisms, e.g., *Pseudomonas putida*, to convert the sugars, usually obtained from corn or other plant residues, into the target product. This is a mixture containing several rhamnolipids, which are released into the liquid phase of the fermentation, while the cells are inactivated and released from the fermentation broth. The final isolation of rhamnolipids during their industrial production consists in the separation and purification of the molecules of interest from the fermentation broth to obtain a pure concentrated aqueous solution containing rhamnolipids [79]. The next few years are expected to see a significant expansion of rhamnolipid production to the multi-million-ton scale. This development is expected to significantly improve the prospects for microbial biosurfactants [22]*.

In addition to the conventional methodologies for producing rhamnolipids through fermentation processes, the production of rhamnolipids using chemical synthesis approaches has gained interest in recent years, especially because this type of strategies can contribute to overcome the complexity of the scaling of the production of rhamnolipids [80, 81]*. However, biotechnological processes continue to be advantageous, than chemical production, in the long run due to reduced material and energy waste [75]*. These synthetic rhamnolipids are produced at commercial scale, and currently marketed as green surfactants. However, their production costs are higher than those of another synthetic surfactant. On the other hand, rhamnolipids produced following synthetic routes are not truly biosurfactants. They should be considered as bio-based surfactant which reduces the consumer perception, even if the used methods are claimed to be green. It should be stressed that synthetic routes allow to produce specific rhamnolipid congeners in a very controlled way [80]. Some of them are very difficult to obtain using biotechnological based approaches [82]**. In fact, Glycosurf offers an extensive catalogue of synthetic rhamnolipids with a wide range of molecular structures, using chemicals from renewable sources as precursors. Its manufacturing methodology allows the production of rhamnolipids with an almost infinite range of molecular structures [83].

**Self-assembly of rhamnolipids in aqueous solutions**

The molecular structure of surfactants leads to a complex phase behavior in aqueous medium due to the ability of surfactants to undergoing self-assembly phenomena to produce, depending on the concentration and temperature, a broad range of different structures, including micelles, vesicles, bilayers and various liquid crystalline mesophases. This is very important for the potential applications of rhamnolipids. Above the so-called critical micelle concentration (CMC), surfactants form micelles which can present different geometries (spherical, disk-like or rod-like) [46, 84].

Wu et al. [85] studied mono- and di-rhamnolipids with the same hydrophobic moieties and found that the CMC values of mono-rhamnolipids were lower than those of di-rhamnolipids. This was explained by the fact that the two rhamnose rings on the polar head of di-rhamnolipids increase the hydrophilic-lipophilic balance (HLB) in relation to mono-rhamnolipids, and therefore for surfactants with the same hydrophobic moieties it can be expected that the higher the number of rhamnose rings on the polar head, the lower the hydrophobic character in agreement with the results by Klosowska-Chomiczewska et al. [86]. They also found that the presence of impurities or mixtures of different rhamnolipids in solution can also induce a strong change in the aggregation pattern. It should be noted that the different hydrophobicity of the rhamnolipids determines the type of aggregates (micelles, vesicles, or lamellae) that are formed above the CMC.

The HLB of rhamnolipids also depends on the nature of the fatty acid tails. This was pointed out by Palos Pacheco et al. [80]. They studied the formation of micelles and their structure for a series of homologous mono-rhamnolipids characterized by the asymmetry of their hydrophobic region. In all cases, one of the chains contained 14 carbons and the second contained a number between 6 and 14 carbons. Their results showed that the difference in the length of one of the hydrophobic chains does not lead to a monotonic dependence on the CMC. This can be rationalized considering the existence of an intricate balance of interactions within the rhamnolipid congeners. This agrees with the conclusions extracted from the molecular dynamics simulations by Euston et al. [87].

The charge of the rhamnolipids can also influence their aggregation pattern in solution, as shown the studies by Munusamy et al. [88] and Eismin et al. [89]. In the former work, Munusamy et al. [88] studied the aggregation of non-ionic rhamnolipids using molecular dynamics simulations and found the formation of a wide range of structures, e.g. spherical, ellipsoidal, toroidal and unilamellar vesicles, while the introduction of charged groups within the hydrophilic head can also drive the formation of long tubular structures [89]. This may be explained considering the different ability to form hydrogen bonds depending on the nature of the specific rhamnolipid considered, and their charge.

The structure of the aggregates formed in solution can be tuned by changing the pH, temperature concentration and presence of electrolytes in the medium [90]. Wu et al. [85] studied the effect of pH and temperature on the self-assembly of mono- rhamnolipid s and di- rhamnolipid s in solution and found that increasing the solution pH promotes a transition from vesicles to micelles. This can be explained by the effect of pH on the degree of dissociation of the rhamnolipids. In fact, the higher the pH, the higher the number of ionized groups in the polar head of the rhamnolipids. This increases the electrostatic repulsions, which leads to a small micelle curvature and drives the transition from large to small aggregates. Therefore, it may be expected that any effect that can modify the ionic equilibrium in rhamnolipid solutions can alter the CMC as well as the shape and size of the aggregates. In fact, the addition of ionic salt to rhamnolipids solutions reduces the CMC driving to the formation of bigger rod-like aggregates as a result of a charge screening phenomenon [91]. The growth of aggregates by salt addition agrees qualitatively with the results obtained by Chen and Lee [92] using dissipative particle dynamics simulations. They found that in the absence of salts (zero-salinity conditions) all the rhamnolipids studied form small ellipsoidal clusters, while increasing the salt concentration leads to the formation of larger aggregates due to the shielding of the electrostatic repulsion between the ionized head. It should be noted that the increase in aggregate size with the addition of salt is independent of the type of salt used [91]. On the other hand, the increase in temperature leads to a transition from micelles to vesicles. This can be rationalized by the fact that the increase in temperature reduces the hydration of the molecules and increases their hydrophobicity. This increases the lateral van der Waals interactions between the hydrophobic parts of the rhamnolipids and forces their organization as bilayers [85].

Finally, the hydrophobicity of the carbon source can also influence the aggregation of rhamnolipids in solution [86]. This can be understood if one considers that the nature of the products obtained depends on the specific nature of the initial substrate, due to the different metabolic pathways that are activated to convert the initial substrate into the final rhamnolipids [93].

**Interfacial properties of rhamnolipids**

The interfacial properties of glycolipids, particularly rhamnolipids, are determined by their molecular structure, that influences their adsorption behavior at fluid interfaces, both air/water and oil/water interfaces. The type and size of the sugar head group, the length and saturation of the hydrophobic moieties, and the presence of any functional groups influence the interfacial properties of rhamnolipids, including their ability to reduce the interfacial tension, which is important for example when rhamnolipids can be used in the bioremediation of oil spills [94]. Furthermore, the ability to reduce the interfacial tension plays a crucial role in the detergency, wetting, emulsifying, solubilizing, dispersing, and foaming effects of rhamnolipids.

The adsorption of glycolipid surfactants to the water/vapor interface results into a strong decrease of the interfacial tension [37, 95], even stronger than that obtained when synthetic surfactants are considered as showed by Khan and Sasmal [96]*. They found that rhamnolipid s were able to reduce the surface tension down to values below to the minimum values obtained for the adsorption of cetyltrimethyl ammonium bromide. In fact, rhamnolipids are, in most cases, very strong anionic surfactants that can reduce the interfacial tension from the value corresponding to a bare water/vapor interface (around 72 mN/m) to values of around 27 mN/m, and that of water/alkane interface down to values close to 1 mN/m [97]. For instance, Rahimi et al. [98] reported that rhamnolipids isolated from *P. aeruginosa* (strain MR01) were able to reduce the water/vapor interfacial tension down to values in the range 29-34 mN/m, with di-rhamnolipids containing two decyl hydrophobic chains provoking a higher decrease in the interfacial tension than the mono-rhamnolipids with the same hydrophobic part. They also found that the CMC slightly increases with the number of rhamnose rings on the polar head. Thus, the CMC of mono-rhamnolipids was found to be around 26 mg/L, while di-rhamnolipids have a CMC of around 30 mg/L. The higher interfacial tension of di-rhamnolipids than mono-rhamnolipids reported by Rahimi et al. [98] contrasts with the results by Wu et al. [85] and by Ikizler et al. [99]. They studied the adsorption of mono-rhamnolipids and di-rhamnolipids with hydrophobic alkyl chains of 10 carbon atoms and found that the higher hydrophobicity of the surfactant with only one rhamnose moiety in its polar head favored its ability to reduce the interfacial tension. These discrepancies are not easy to rationalize, although they may be due to differences in purity. This parameter is only reported in the work of Ikizler et al. [99], where rhamnolipids with purity above 95% were used. Based on the above results, it can be assumed that the definition of the interfacial activity of rhamnolipids is not trivial and requires an analysis of the purity of the rhamnolipid samples. This is important because rhamnolipids are generally obtained as complex mixtures containing different congeners. Wu et al. [85] also reported that regardless of the number of rhamnose units on the polar head, the stability of both rhamnolipids to pH and temperature changes appears to be very high. The analysis of the dynamic interfacial tension showed that during the first steps of adsorption, the process is diffusion controlled independently of the type of rhamnolipid considered, with the effective diffusion of mono-rhamnolipids being higher than that corresponding to di-rhamnolipids.

The adsorption of rhamnolipids at the water-vapor interface is strongly dependent on the specific structure. In general, the maximum excess surface concentration is 40% higher for the adsorption of mono-rhamnolipids than for the adsorption of di-rhamnolipids (60), which can be explained by the larger size of the polar headgroups in rhamnolipids, i.e., di-rhamnolipids contain two rhamnose units in the polar head, whereas mono-rhamnolipids have only one. This increases the steric hindrance to adsorption and therefore reduces the excess surface concentration. This is particularly important when considering the adsorption of mixtures containing mono-rhamnolipids and di-rhamnolipids. In this case, the water-vapor interface is enriched in mono-rhamnolipids which is consistent with their higher hydrophobicity [100].

The addition of ionic salts to rhamnolipids solutions increases the ability of the latter to decrease the interfacial tension of the water/vapor interfaces. This can be explained considering that salt ions interact with the carboxylate ions in the polar headgroup of the rhamnolipids, leading to a charge screening phenomenon. Thus, the intermolecular electrostatic repulsions are reduced and the molecules can pack more closely at the interface [91].

The specific carbon source used to produce rhamnolipids also influences their interfacial behavior. In fact, rhamnolipids obtained using glucose or soybean oil as carbon source produces mixtures of rhamnolipids with different composition. These mixtures reduce differently the interfacial tension, with the rhamnolipids obtained from glucose leading to a stronger decrease in the interfacial tension than those obtained using soybean oil as carbon source. This is explained by the higher contain of mono-rhamnolipids in the mixtures obtained using glucose as the carbon source [101].

**Emulsifying properties of rhamnolipids**

The ability for emulsifying oily compounds is associated with the mono-rhamnolipid content in the mixture produced by specific microorganisms. In fact, Zhao et al. [102] compared the ability of rhamnolipids produced by different *Pseudomonas* strains for recovering oil from oily sludge, and found that this is enhanced as the concentration of rhamnolipids in the produced mixtures is increased. This is explained considering that the less hydrophilicity of mono-rhamnolipids in relation to di-rhamnolipids contributes to enhance their emulsifying activity. Similar results were reported by Rocha et al. [103]. They used rhamnolipids produced by different bacteria strains to emulsifying different petroleum fraction and hexadecane and found that the use of rhamnolipids mixtures containing mostly mono-rhamnolipids leads to a very efficient emulsification independently of the oil nature, with the emulsification being enhanced as the content of mono-rhamnolipids is increased. On the other hand, when di-rhamnolipids are the main component of the mixture, the emulsification ability is significantly worsened. In fact, mono-rhamnolipids leads to the emulsification of the whole liquid volume, while the use of di-rhamnolipids reduce the emulsified volume down to 65% of the total volume. This better performance of mono-rhamnolipids in emulsion stabilization contrasts with the results by Zhao et al. [101] in a more recent study. They reported that mixtures of rhamnolipids obtained with different carbon sources and containing a higher proportion of di-rhamnolipids improved the emulsification process. However, the reason for the improved emulsification in the mixture containing a higher fraction of di-rhamnolipids is not due to the di-rhamnolipids themselves, but to the higher hydrophobicity of the alkyl moieties. Thus, the higher the hydrophobicity of the carbon source, the higher the hydrophobicity of the rhamnolipids obtained and their better emulsifying capacity. It should be noted that according to the work by Li et al. [104], the effects of the rhamnolipid nature on the emulsification ability is suppressed when the concentration of surfactant is very high, in the range of the one used in industrial applications. This was confirmed by Li et al. [104] They studied the emulsification of xylene and cyclohexane using rhamnolipid mixtures containing several congeners at a concentration of 10 g/L and found that at such a high concentration, the composition of the emulsifying mixture does not significantly affect the ability to emulsify the oil droplets and the stability of the emulsions obtained.

Al-Shakkaf and Onaizi [105] have recently investigated the ability of rhamnolipids to stabilize crude oil-in-water nanoemulsions at different pH and salinity levels. They found that rhamnolipids are a very promising alternative to stabilize nanoemulsions under highly alkaline conditions (pH around 12), leading to the formation of droplets with an average size of around 38 nm. Under such conditions, the nanoemulsions obtained exhibit negative zeta potential and long-term stability. The addition of salt contributes to the destabilization processes of the obtained dispersion due to charge screening. This leads to the complete destabilization of the nanoemulsions within 48 hours when the ionic strength is remarkably high. Similarly, the reduction in pH contributes to the destabilization by reducing the effective charge of the rhamnolipids molecules, forcing droplet coalescence and Ostwald ripening. In fact, the addition of 0.5 M HCl leads to a rapid destabilization, resulting in complete demulsification after 1 h, regardless of the initial pH or ionic strength. This is explained by the high tendency of rhamnolipids to aggregate at very low pH. It should be noted that destabilization of the emulsions is also possible by adding NaOH when the salinity of the dispersion is relatively high. This can be understood by the fact that although the addition of NaOH increases the deprotonation of the rhamnolipid molecules, which should increase the repulsion between the oil droplets, it also increases the Na+ content and therefore leads to a prevalence of van der

Waals attractive forces resulting from screening phenomena, which can counteract the deprotonation effect. This leads to the destabilization of the nanoemulsions.

**Applications of rhamnolipids**

The range of applications for rhamnolipids is enormous [106-109]. For example, their ability to reduce interfacial tension has played a very important role in cosmetics, detergents and other industries producing washing products [110]. In addition, rhamnolipids have interesting antimicrobial, anticancer or immunomodulatory properties [111-115]. Table 3 summarizes their potential applications [37].

Table 3. Example of some of the potential applications of rhamnolipids. Adapted from Salek et al. [37]. Licensed under Creative Common Attribution 4.0 International License.

| Biological or physicochemical activity | Application Field | Reference |
|---|---|---|
| Antimicrobial | Food science | [113] |
| Cytotoxic effect on human breast cancer cells | Medical | [98] |
| Stabilization of oil in high water internal phase emulsions | Food Science Medical Cosmetic and personal care | [107] |
| Anti-bacterial activity towards food pathogens: *Bacillus cereus*, *Listeria Monocytogenes* and *Staphylococcus aureus* | Food packaging | [111] |
| Nanoemulsions for drug delivery mechanism against SCC7 tumor cells | Medical | [114] |
| Biodegradation of hydrophobic organic compounds | Bioremediation | [108] |
| Microbial-enhanced oil recovery | Environmental protection Petroleum industry | [109] |
| Nanoparticle synthesis | Medical | [115] |

**Biomedicine and biotechnology applications.** The environmentally friendly and biodegradable nature of rhamnolipids and their multiple biological properties have stimulated research into their use as a therapeutic option against various diseases [18]. However, the practical application of rhamnolipids for clinical use requires lengthy and expensive regulatory approval, and therefore the transition from laboratory to clinical scale is not easy in the case of rhamnolipids.

One of the most promising biomedical applications of rhamnolipids is related to their potential role in anti-tumoral therapy. rhamnolipids can inhibit the production of breast cancer cells (MCF-7 line) [116]. In addition, rhamnolipids have wound-healing activity, recognize the cytoskeleton of phagocytic/nonphagocytic cells and alter their morphology [117]. This morphological change is a signature of apoptosis. Moreover, rhamnolipids can reduce cell viability through their interaction with cell membranes [98]. On the other hand, mono-rhamnolipids are more potent antitumor agents than di-rhamnolipids. In fact, the dose required to induce a cytotoxic effect is lower than that required for di-rhamnolipids. Mono-rhamnolipids can induce changes in the morphology of leukemic cells, chromatic condensation, and the appearance of apoptotic bodies and nuclear fragmentation, which ultimately leads to cell apoptosis. In addition, the use of higher concentrations of mono-rhamnolipids can lead to the overexpression of specific genes that induce rapid proliferation and anti-apoptotic behavior [117].

Rhamnolipids can also be used as a tool for the control of post-harvest damage in various crops [118]. The use of rhamnolipids as an alternative to chemical fungicides has grown significantly in recent years, taking advantage of their antifungal activity against various fungi, such as *Alternaria alternata*, *Mucor circinelloides* and *Verticillium dahlia*. In addition, the use of rhamnolipids avoids the phenomena of fungicide resistance and helps to stimulate plant immunity against pathogens [118, 119]. Rhamnolipids can inhibit fungal growth, and therefore a certain concentration is required to maximize their effect. However, if the concentration exceeds a threshold value, the growth of the fungus is hindered but not inhibited. In addition, at extremely high concentrations, rhamnolipids can cause cell damage, reducing their ability to fight the fungal infection [120]. Rhamnolipids can also induce a remodeling of the cell membrane of *Alternaria alternata*, together with a change in hyphal morphology, which inhibits mycelial growth and spore germination of the fungus, thus hindering fungal growth [118]. Similar effects have been reported against other fungus families [18]. For instance, Sen et al. [121] conducted further studies on the antifungal activity of rhamnolipids, specifically focusing on their impact on dermatophytosis induced by *Trichophyton rubrum* in mice models. Their research revealed that rhamnolipids effectively inhibited spore germination and hyphal proliferation in *Trichophyton rubrum*. Moreover, fungal cells treated with rhamnolipids exhibited significant alterations in the hyphal region and a release of nucleic acids due to compromised cell membrane integrity. Importantly, the topical application of rhamnolipids at a concentration of approximately 500 µg/mL proved highly effective in curing dermatophytosis over a 21-day treatment period. These findings are comparable to the results obtained with conventional antifungal drugs such as terbinafine. This opens interesting opportunities for the use of rhamnolipids in the treatment of dermatophytic infections. Recently, it has been reported that the stronger antifungal effect in rhamnolipids is found for di-rhamnolipids. Unfortunately, the action mechanism remains unclear yet [104].

Recently, the potential role of rhamnolipids as immunomodulators has also been explored. Indeed, rhamnolipids can contribute to the modulation of the immune system, both humoral and cellular, can contribute to the activation of immune cells and induce the secretion of pro-inflammatory cytokines [122]. On the other hand, preincubation of rhamnolipids with monocytes can induce an increase in cellular oxidative stress. In addition, rhamnolipids can contribute to the induction of histamine, serotonin and 12-hydroxyeicosatetraenoic acid production in certain cells. Rhamnolipids has also been shown to be effective in reducing the activity of macrophages by inducing their lysis. In the case of polymorphonuclear leukocytes, rhamnolipids can induce necrosis [120, 123].

Biofilm removal is another very important application of rhamnolipids [124]. This is possible due to the ability of rhamnolipids to modulate various forces associated with biofilm deposition, including capillary forces, contact angle and interfacial tension [125]. It should be noted that the role of rhamnolipids in biofilm removal is strongly dependent on two factors: (1) the composition of the biofilm matrix and (2) the culture medium. This is crucial when considering biofilm removal in food. In fact, the presence of carbohydrates in the medium facilitates the removal of biofilm due to the possible interactions between carbohydrates and rhamnolipids [126]. On the other hand, coating surfaces with a layer formed by rhamnolipids can reduce the ability of biofilm formation, as demonstrated by Tambone et al [127]. by depositing rhamnolipids on the surface of dental titanium implants. The ability of rhamnolipids to remove biofilms was also highlighted by Kim et al. [128]. They showed that the use of a rhamnolipid solution with a concentration around the CMC against a biofilm of *P. aeruginosa* allowed the removal of negatively charged humic-like, protein-like and fulvic acid-like substances. rhamnolipids also reduce the concentration of extracellular polymeric substances such as carbohydrates and proteins, and therefore rhamnolipids play a very important role in combating biofilm formation. It should be noted that the ability of rhamnolipids to disrupt biofilm is nutrient specific and dependent on the matrix composition of the biofilm. In fact, the activity of rhamnolipids in the disruption of biofilms can be hindered by the specific nature of the culture medium as pointed out the studies by Silva et al. [125].

The antimicrobial role of rhamnolipids has also been demonstrated by Gaur et al. [41]. They showed that rhamnolipids isolated from various sources have bactericidal activity against pathogenic bacteria, both gram-positive and gram-negative. This is possible because rhamnolipids can increase the permeability of the bacterial cell membrane and distort its structure, leading to increased release of DNA and proteins into the extracellular space. In addition, rhamnolipids can generate reactive oxygen species that help to kill bacteria. The results by de Freitas Ferreira et al. [111] were consistent with the above scenario, and explain that the sensitivity of different microorganisms to rhamnolipids is related to the ability of these molecules to reduce the hydrophobicity of the cell surface, and induce damage in the cytoplasmic membrane. This may explain the different effects of rhamnolipids against different microorganisms. For example, rhamnolipids can contribute to the removal of gram-positive bacteria, especially under acidic conditions. However, gram-negative bacteria were found to be insensitive to rhamnolipids.

An additional biomedical application of rhamnolipids is associated with their wound-healing ability. The application of rhamnolipid extract produced by *Acinetobacter calcoaceticus* can stimulate the mouse fibroblast L929 cells. This is associated with the ability of rhamnolipids to enhance Smad3 phosphorylation in L929 which plays a significant role for controlling fibroblast migration. Therefore, rhamnolipids contribute to promote wound healing in mice with excisional wound by increasing the protein levels of TGF-$\beta$1 and alpha smooth muscle actin [44].

The above discussion, although not exhaustive, has shown that the various physico-chemical and biological properties of rhamnolipids make them a powerful tool for their application in the prevention of various problems affecting human health.

**Applications in agriculture.** The potential of rhamnolipids in the chemical control of pathogens of interest to agriculture has been known for more than two decades [129]. This use is generally associated with the antimicrobial properties of rhamnolipids, which allow the control of various pests and improve the uptake of nutrients by plants [12]. In addition, rhamnolipids can also be used to facilitate the association between plants and micro-organisms, to enhance plant growth and to improve the quality of the soil [130]. Zhao et al. [131] proposed that the anti-microbial properties of rhamnolipids can be exploited to develop green pesticides for agricultural applications. In particular, mono-rhamnolipids emerge as a very promising alternative for this aim due to the relatively low dose required for producing strong effects against pests. Beyond the applications as pesticides, rhamnolipids can find another uses in agriculture. Ren et al. [132] found that rhamnolipids can contribute to composting. In fact, the addition of rhamnolipids after Fenton pretreatment and inoculation of fungi into the compost helps to the degradation process of the organic matter and the formation of humic substance. This is possible because rhamnolipids promoted the formation of lignocellulose-degrading products. On the other hand, the addition of rhamnolipids to soils improves soil stability, lowers pH, and produces salt rejection, helping to alleviate salt stress on microorganisms and plants. In addition, rhamnolipids help to improve microbial growth and activity as well as germination performance. rhamnolipids therefore modify the microbial communities in the soil and support the proliferation of bacteria. This improves soil properties and nutrient cycling [133].

**Applications in enhanced oil recovery and bioremediation.** The use of rhamnolipids can enhance significantly the amount of oil recovered [134]. rhamnolipids can be also exploited to remove crude oil, heavy metals, and other toxic compounds from contaminated substrates both soils and water [135, 136].

The use of biosurfactants, and in particular rhamnolipids, for bioremediation purposes is related to the specificity of microorganisms to utilize organic and hydrocarbon wastes as raw materials. Furthermore, biosurfactants have a higher surface activity than conventional surfactants and are more resistant to environmental factors, including extreme conditions of acidity or basicity of aqueous solutions, temperature, or salt concentration. On the other hand, biodegradability and demulsifying/emulsifying ability are also of paramount importance in the application of biosurfactants for bioremediation purposes [137]. In fact, the performance of rhamnolipids in remediation processes is commonly associated with their role on the emulsification or solubilization of hydrocarbons and their ability to modify bacterial cell surface properties to enhance interfacial uptake of hydrocarbons [33]. This has been demonstrated by Bhosale et al. [138]. They found that iron oxide nanoparticles capped with rhamnolipid molecules used as a photocatalyst with sodium dodecyl sulphate as an adsorbent allowed almost complete decolorization of methyl violet.

Olasanmi and Thring [139] investigated the ability of rhamnolipids to wash drill cuttings and oil-contaminated soils. They found that the optimum rhamnolipid concentration and washing time were 500 mg/L and 30 min, respectively. Using these conditions, the maximum reduction of total petroleum hydrocarbons from drill cuttings and contaminated soils was approximately 75% and 60%, respectively. Gaur et al. [42] showed that rhamnolipids isolated from *Lysinibacillus sphaericus* at a concentration of 90 mg/L can assist in the dissolution of hydrophobic pesticides. In fact, the obtained rhamnolipids enhanced the ability to dissolve α-, β-endosulfan and γ-hexachlorocyclohexane up to 7.2, 2.9 and 1.8 times, respectively, when compared to the ability of synthetic surfactants such as Triton X-100. On the other hand, the obtained rhamnolipids use benzoic acid, chlorobenzene, 3- and 4-chlorobenzoic acid as carbon sources and have an increased resistance to various heavy metals such as arsenic, lead and cadmium. Gaur et al. [41] demonstrated that different bacteria (*Planococcus rifietoensis* and *Planococcus halotolerans*) can help on the decontamination of pesticide contaminated soils by using the pollutants as carbon source for the production of rhamnolipids. This microbial enhanced decontamination can also be exploited for oil recovery as was demonstrated by

Elakkiya et al. [140]. They produced rhamnolipids (0.34 mg/mL) using cassava solid waste as carbon source in presence of *P. aeruginosa*. The ability of rhamnolipids for oil recovery was also reported by Dai et al. [141]. They found that the combination of rhamnolipids with slow-release nutrients can contribute to the biodegradation of heavy oils present in a contaminated intertidal zone. This is possible because rhamnolipids increases the metabolic activity of the microorganisms involved in heavy oil degradation, leading to a 2-fold increase in degradation yield compared to natural attenuation phenomena. In addition, the presence of rhamnolipids contributes to the simultaneous degradation of n-alkanes and polycyclic aromatic hydrocarbons.

Patowari et al. [142] showed that *P. aeruginosa* (strain SR17) could grow in culture media containing naphthalene as sole carbon source. In fact, the growth of *P. aeruginosa* was found to occur in media with naphthalene concentrations ranging from 0.1% to 0.8%, leading to an optimal growth and rhamnolipid production when the concentration of the aromatic compounds is around 0.3%. In addition, higher naphthalene concentrations reduce the production yield due to the toxicity imparted to the bacteria by naphthalene. The ability of the bacteria for degrading the naphthalene and producing rhamnolipids achieved is maximum after five days, arriving to a maximum percentage of degradation of around 89%. On the other hand, it was found that the produced biosurfactant presents a high ability for emulsifying hydrophobic compounds, which can contribute to the degradation of oily pollutants dispersed in water. These results open interesting perspective for the utilization of rhamnolipid-producing bacterial for the bioremediation of polyaromatic hydrocarbons (PAHs) or oil-contaminated sites. However, it is needed to optimize the culture condition to obtain a more feasible and economical degradation.

Barrios San Martin et al. [143] investigated the ability of rhamnolipids as soil washing agents for the remediation of heavy metal contaminated sediments and found that the use of solutions containing rhamnolipids is a very effective alternative for the removal of heavy metals such as vanadium. In particular, the treatment for 72 h with a solution of 240 mg/L concentration allows the maximum efficiency in vanadium removal (85.5% of vanadium is removed). McCawley et al. [82]* showed that the ability of rhamnolipids to remove metals from natural and contaminated aqueous systems is strongly dependent on the hydrophobicity of the rhamnolipids. In fact, the higher the hydrophobicity, the higher the metal recovery yield. Furthermore, the use of rhamnolipids with tailored structure was shown to be a promising strategy for the selective removal of metals of different nature.

**Applications in cosmetics.** Rhamnolipids have been used in different cosmetic formulations such as acne pads, anti-dandruff products, deodorants, nail care products and toothpastes [95]. In addition, they use in the manufacture of shampoos has recently been explored [144-148]. Indeed, rhamnolipid can replace traditional sulfate-based surfactant, such as sodium dodecyl sulfate or sodium laureth sulfate from hair care and conditioning products [149]. Most of the above products require surfactants with high surface and emulsifying activity. On the other hand, cosmetics containing rhamnolipid have been patented as anti-wrinkle and anti-ageing products and have been commercialized in various dosages [95]. In addition, rhamnolipid can also be used as non-toxic bio-preservative in personal care and cosmetic formulations due to their ability to inhibit the proliferation of different fungi and bacteria [150].

**Concluding remarks**

Rhamnolipids have emerged as a promising class of biosurfactants with a wide range of applications and significant potential for sustainable industrial processes. Rhamnolipids are gaining attention because they can be produced using safe microorganisms and naturally occurring sustainable resources. This not only ensures a renewable supply, but also reduces the environmental impact associated with traditional petrochemical-based surfactant production methods. In addition, the beneficial properties of rhamnolipids increase their appeal as alternatives to petrochemical surfactants and additives in various industries. Their non-toxic nature ensures minimal harm to human health and the environment. rhamnolipids are also highly biodegradable, allowing them to break down into harmless by-products, reducing pollution and promoting environmental sustainability. In addition, rhamnolipids exhibit exceptional surfactant properties, reducing surface tension and achieving low critical micelle concentration (CMC), which enhances their effectiveness in emulsifying and dispersing hydrophobic compounds.

Beyond their surfactant capabilities, rhamnolipids have a wide range of properties that broaden their potential applications. Their antimicrobial, antitumor, antibiofilm and antifungal activities make them valuable in areas beyond traditional detergents, such as pharmaceuticals, healthcare, and agriculture. This versatility positions rhamnolipids as multi-functional agents capable of addressing different challenges in different industries. By harnessing the benefits of rhamnolipids, industries can move towards a more sustainable and environmentally friendly future without compromising product performance and functionality. Integrating rhamnolipids into existing processes can lead to reduced reliance on petrochemicals and mitigate the negative impacts associated with their extraction and production. In addition, rhamnolipids contribute to the development of environmentally friendly solutions that meet the growing demand for sustainable practices. Continued research and development in the field of rhamnolipids will play a crucial role in exploiting their full potential. Indeed, further investigation into optimizing production methods, exploring new applications, and understanding their mechanisms of action will drive the widespread adoption of rhamnolipids in various sectors. This continued commitment to research will enable industry to fully exploit the benefits of rhamnolipids and pave the way for a more sustainable and environmentally conscious future.

**Acknowledgements**

This work was partly funded by MINECO (Spain) under grant PID2019-106557GB-C21, and by the E.U. in the framework of the European Innovative Training Network-Marie Sklodowska-Curie Action NanoPaInt (grant agreement 955612). The graphical abstract has been designed using assets from Freepik.com (https://www.freepik.com).

**Declaration of interest**

The authors declare no conflict of interest.